\documentclass[12pt]{iopart}

\usepackage{graphicx}  
\begin{document}

\title[Hydrodynamic analysis of heavy ion collisions at RHIC]{Hydrodynamic analysis of heavy ion collisions at RHIC}

\author{T Hirano}

\address{
Department of Physics, the University of Tokyo,
Tokyo 113-0033, Japan
}
\ead{hirano@phys.s.u-tokyo.ac.jp}
\begin{abstract}
Current status of dynamical modeling of 
relativistic heavy ion collisions and
hydrodynamic description of
the quark gluon plasma 
is reported.
We find the hadronic rescattering effect plays
an important role in interpretation of 
mass splitting pattern in the differential
elliptic flow data observed at RHIC.
To demonstrate this, 
we predict the elliptic flow parameter
for $\phi$ mesons to directly observe
the flow just after hadronisation.
We also discuss recent applications of outputs from
hydrodynamic calculations to $J/\psi$ suppression, 
thermal photon radiation and heavy quark diffusion.
\end{abstract}


\section{Introduction}

Understanding of the thermodynamic and transport properties of
the quark gluon plasma (QGP) is
one of the main topics in the physics of 
relativistic heavy ion collisions \cite{YagiHatsudaMiake}.
The QGP created in these reactions
is by no means a static matter
and exists as a transient state
whose lifetime is of the order of $\sim$ 1-10 fm/$c$.
It is non-trivial even whether local equilibrium
is really reached or not,
which should be also clarified through the observation.
Hence, one first has to justify the local equilibration
by comparing the model calculation
with experimental data.
This is the starting point of further studies
on the QGP under equilibrium.
In order to understand thermodynamical (bulk) 
and transport properties of the QGP
in relativistic heavy ion collisions,
one then has to develop a \textit{dynamical} 
analysis tool which links the observables
with the equation of state (EoS)
and transport coefficients of the QGP.
It is inevitable to apply each relevant model
to each stage of the collisions
and unify these models consistently to describe the whole stage
of the reactions.

Along the line of thought, we have constructed a dynamical and unified model 
based on a
fully three-dimensional
hydrodynamic description of the QGP \cite{Hirano:2005xf,Hirano:2007ei}.
One of the major discoveries 
at relativistic heavy ion collider (RHIC)
is that
elliptic flow parameters, which are intimately related with
the degree of thermalisation,
are well described
within a framework of relativistic hydrodynamics \cite{Hirano:2008hy}.
An assumption of local thermalisation 
has to be made in hydrodynamics.
However, local thermalisation can
be expected only in the intermediate stage.
Thus, one needs to model
dynamics of reactions appropriately 
before and after the hydrodynamic stage
to draw informations of the QGP from experimental data.
Once we obtain the hydrodynamic evolution of the QGP,
information of 
local thermodynamic variables is available
to analyse other observables such as 
$J/\psi$ suppression, heavy quark diffusion and
thermal radiation.
In this way, hydrodynamic models
enable one to analyse a vast body of data 
in relativistic heavy ion collisions.

\section{Dynamical modeling of heavy ion collisions}

Hydrodynamics
is a general framework to describe the space-time evolution
of locally thermalised matter for a given EoS.
This framework has been applied to the intermediate stage in 
heavy ion collisions.
We neglect the effects of dissipation
and concentrate on discussion about ideal hydrodynamic models.
The main ingredient in ideal hydrodynamic models
is the EoS
of hot and dense matter governed by QCD.
In addition, one also needs to assign initial conditions
to the hydrodynamic equations.
Hydrodynamics can be applied to a
system as long as its local thermalisation is maintained.
However, in the final
stage the particles are freely streaming toward the detectors and
their mean free path is almost infinite.
This is obviously beyond the applicability of hydrodynamics. 
Hence we also need a 
description to decouple the particles from the rest of the system.
The hydrodynamic modeling of relativistic heavy 
ion collisions needs an EoS, initial
conditions and a decoupling prescription. 
Hydrodynamic modeling
has been sophisticated for these years and tested against 
RHIC data \cite{Hirano:2008hy}.

 The EoS is in principle calculated from lattice QCD simulations.
Instead of using the result from lattice QCD simulations,
we employ a phenomenological model in this study.
At high energy density, the EoS can be described by a bag model
\begin{equation}
P = \frac{1}{3}(e-4B)\,.
\end{equation}
Here $P$ is pressure and $e$ is energy density.
The bag constant $B$ is tuned to 
match pressure of the QGP phase to that of a hadron resonance gas
at critical temperature $T_c$: $P_{\mathrm{QGP}}(T_c) = P_{\mathrm{hadron}}(T_c)$.
A hadron gas in heavy ion collisions
is not in chemical equilibrium below the chemical freezeout
temperature $T^{\mathrm{ch}}$ which is close to $T_c$
\cite{BraunMunzinger:2003zd}, so the hadron phase
may not be chemically equilibrated.
A chemically frozen (or only partially equilibrated)
hadron resonance gas can be described
by introducing the chemical potential for each hadron \cite{Bebie:1991ij,
Hirano:2002ds}.

For a decoupling prescription,
the Cooper-Frye formula \cite{Cooper:1974mv} is almost a unique choice to convert the hydrodynamic
picture to the particle picture.
Contribution from resonance decays should be taken into
account by applying decay kinematics to the outcome of the
Cooper-Frye formula.
The decoupling temperature $T^{\mathrm{dec}}$
is fixed through \textit{simultaneous} fitting of
$p_{T}$ spectra for various hadrons in the low $p_{T}$ region.

The prescription to calculate the momentum distribution as above
is sometimes called the sudden freezeout model
since the mean free path of the particles changes
from zero (ideal fluid) to infinity (free streaming) 
within a thin layer.
Although this model is too simple, it has been used in hydrodynamic
calculations for a long time.
Recently one utilises hadronic cascade models to describe
the gradual freezeout
\cite{Hirano:2005xf,Hirano:2007ei,Bass:2000ib,Teaney:2000cw,Nonaka:2006yn}
in a more realistic way.
As will be shown, the hadronic afterburner is mandatory 
in understanding
elliptic flow data.
Phase space distributions for hadrons
are initialized below $T_c$ by using the Cooper-Frye formula.
We employ a hadronic cascade model JAM \cite{Nara:1999dz} to describe
the space-time evolution of the interacting hadron gas.
This kind of hybrid approaches in which the QGP fluids are 
followed by hadronic cascade models
automatically describes both the chemical and
thermal freezeout and is much more realistic especially 
for the late stage than the naive Cooper-Frye prescription.

Initial conditions in hydrodynamic simulations
are so chosen as to reproduce the
centrality and rapidity dependences of multiplicity $dN_{\mathrm{ch}}/d\eta$.
Initial
conditions here 
are
energy density distribution $e(x,y,\eta_s)$ and flow velocity
$u^{\mu}(x,y,\eta_s)$ at the initial time $\tau_0$.
Again baryon density is neglected since the net baryon
density is quite small at midrapidity at RHIC.
At mid-(space-time)rapidity $\eta_s = 0$, one can parametrise the initial entropy density based
on the Glauber model
\begin{equation}
s(x,y) = \frac{dS}{\tau_0 d\eta_s d^2x_\perp}  \propto 
\alpha n_{\rm part}(x,y; b) 
+ (1{-}\alpha)n_{\rm coll}(x,y; b)
\label{eq:entroini}
\end{equation}
The soft/hard fraction $\alpha$ is adjusted to reproduce the 
measured centrality dependence \cite{PHOBOS_Nch} of the charged hadron 
multiplicity at midrapidity.
By using the EoS, one can calculate the initial energy density distribution
together with pressure and temperature distributions
from Eq.~(\ref{eq:entroini}).
For space-time rapidity dependence of initial conditions
and a novel initial condition
based on the colour glass condensate (CGC) picture,
see Refs.~\cite{Hirano:2005xf,Hirano:2004rs}.
The fitting of multiplicity is the starting point of further analysis based on 
hydrodynamic simulations.

In the hydrodynamic models, various combinations of initial
conditions, EoS and decoupling prescriptions
are available to analyse the experimental data.
Of course, final results largely depend on modeling of each stage.
So it is quite important to constrain each model and its inherent parameters
through systematic analysis of the data toward a comprehensive understanding 
of the QGP.

\section{Elliptic flow}

Anisotropic transverse flow can be quantified by
Fourier components of azimuthal distribution,
$v_n = \langle \cos n \varphi \rangle$,
where the average is taken over by weighting
azimuthal distribution.
The second harmonics $v_2$ is called elliptic flow parameter
and is intimately related with the EoS and
the degree of thermalisation \cite{Ollitrault:1992bk}.

\subsection{Charged hadrons}

\begin{figure}
\centering
\includegraphics[width=0.46\textwidth]{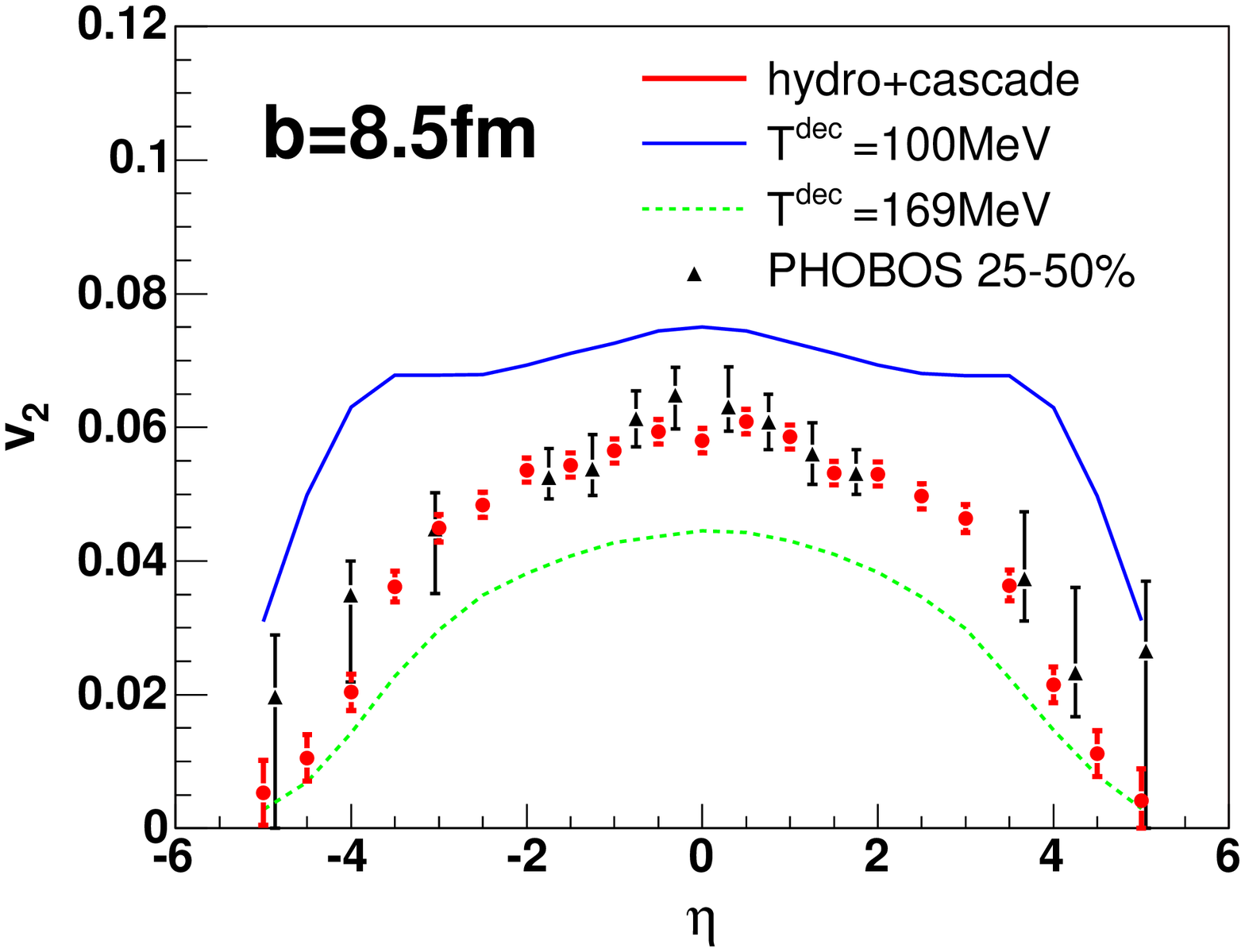}
\includegraphics[width=0.46\textwidth]{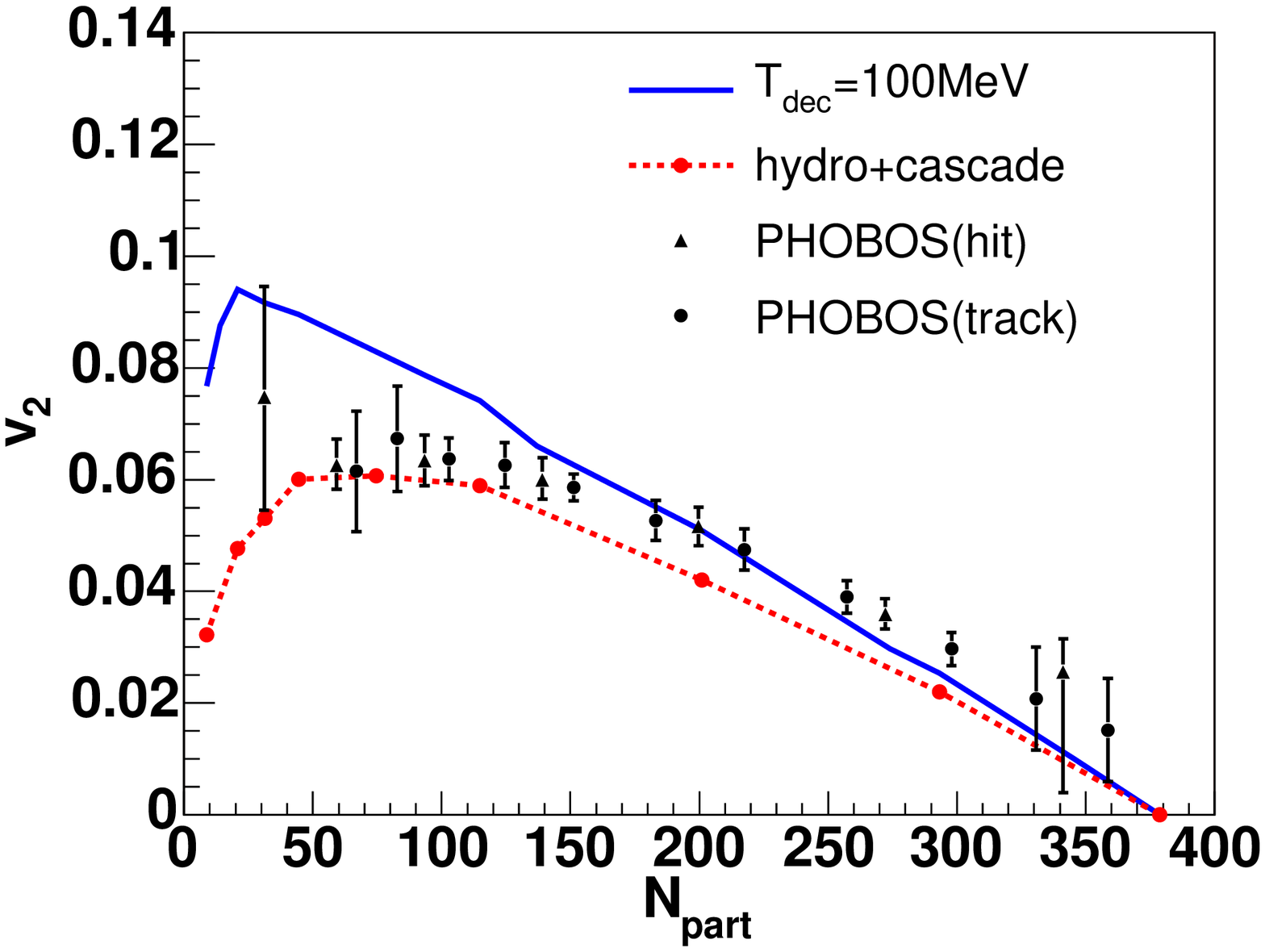}
\caption{(Left) Pseudorapidity dependence
of $v_2$ for charged hadrons.
PHOBOS data \cite{Back:2004mh} are compared to 
the ideal hydrodynamic model with
$T^{\mathrm{dec}}=100$ and 169 MeV
and the hybrid model \cite{Hirano:2005xf}.
(Right) $v_2$ for charged hadrons as a function of centrality
in $\mid \eta \mid < 1$. PHOBOS data \cite{Back:2004mh} are compared to 
the ideal hydrodynamic model with $T^{\mathrm{dec}}=100$ MeV
and the hybrid model \cite{Hirano:2005xf}.
}
\label{fig:v2charged}
\end{figure}

The pseudorapidity dependence of $v_2$
observed by PHOBOS \cite{Back:2004mh} has a triangular shape:
$v_2$ is maximum at midrapidity and decreases
as moving away from midrapidity.
On the other hand, in the ideal hydrodynamic calculations
with $T^{\mathrm{dec}}=100$ MeV,
$v_2$ depends mildly on pseudorapidity in $\mid \eta \mid < 3$.
Consequently, the hydrodynamic model
generates as large $v_2$ as
the data only near midrapidity
and overshoots the data significantly
at forward and
backward rapidities \cite{Hirano:2002ds,Hirano:2001eu}
 as shown in Fig.~\ref{fig:v2charged} (left).
If we replace the
hadron fluid with a hadron gas utilising a hadronic cascade,
$v_2$ is generated little
in the forward and backward region during hadronic
rescattering stage \cite{Hirano:2005xf}.
In this hybrid model the hadrons have a finite mean free path
and exhibit an effective shear viscosity in the hadron phase.
So the dissipative hadronic ``corona" effect \cite{Hirano:2005wx}
turns out to be important in understanding
the charged hadron $v_2$ data.
Figure~\ref{fig:v2charged} (right) shows the centrality dependence of $v_2$
at midrapidity ($\mid \eta \mid < 1$).
The solid line is the result from an ideal hydrodynamic calculation, 
while the dotted line from the
hybrid model.
It is clear that for peripheral collisions where the multiplicity is small
the hadronic viscosity plays an important role in 
description of the $v_2$ data.
One may notice that the result from the hybrid model
systematically and slightly smaller
than the data. However, there could exist an effect of initial eccentricity
fluctuations which is absent in the present hydrodynamic calculations
and could enhance $v_2$ by $\sim90\%$
in 0-5\% and by $\sim 20\%$ in 50-60\%
centralities \cite{HiranoNara_unpublished}.
The deviation between the results 
and the data may be interpreted quantitatively
by this eccentricity fluctuation effect.

\subsection{Identified hadrons}

\begin{figure}
\centering
\includegraphics[width=0.46\textwidth]{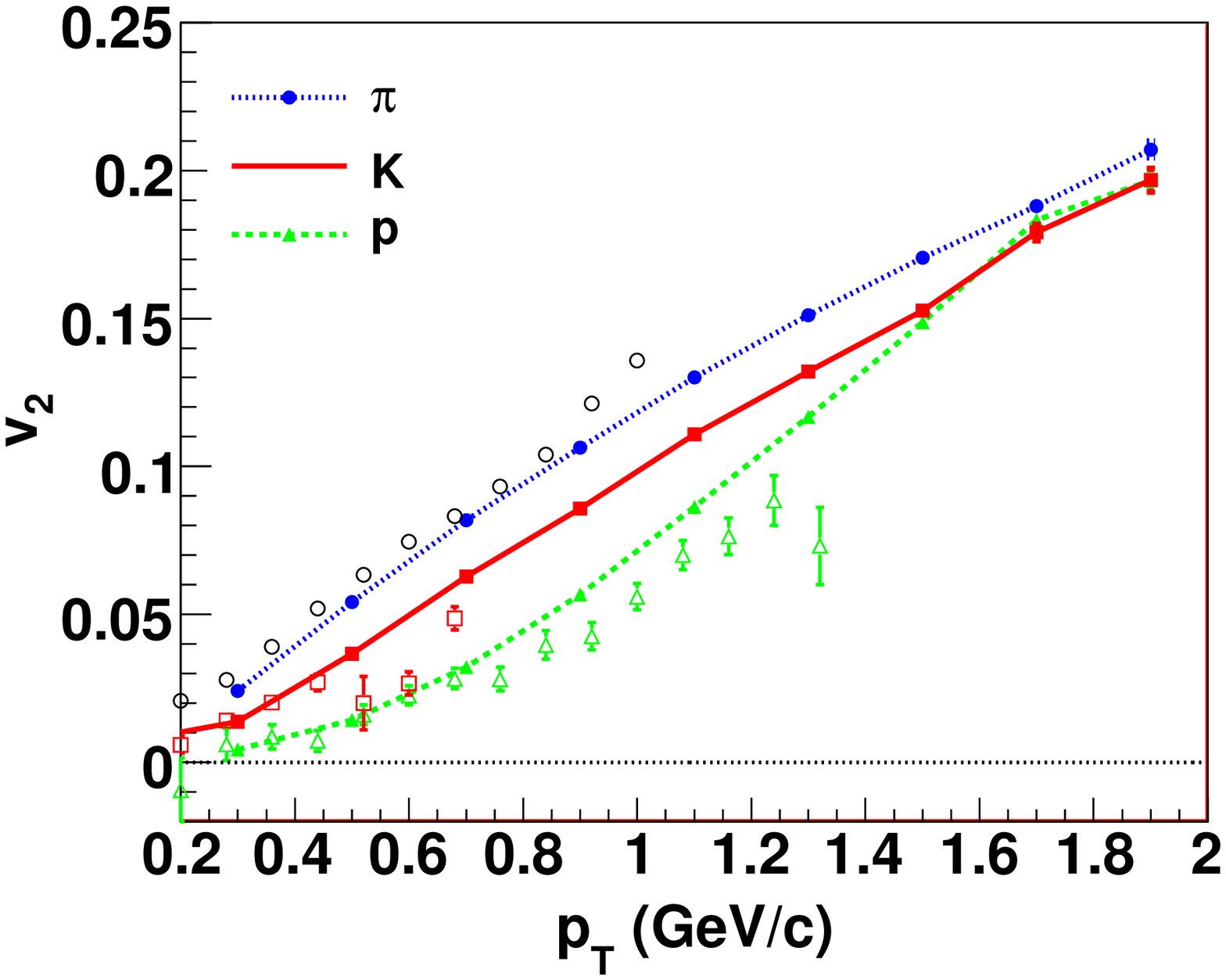}
\includegraphics[width=0.46\textwidth]{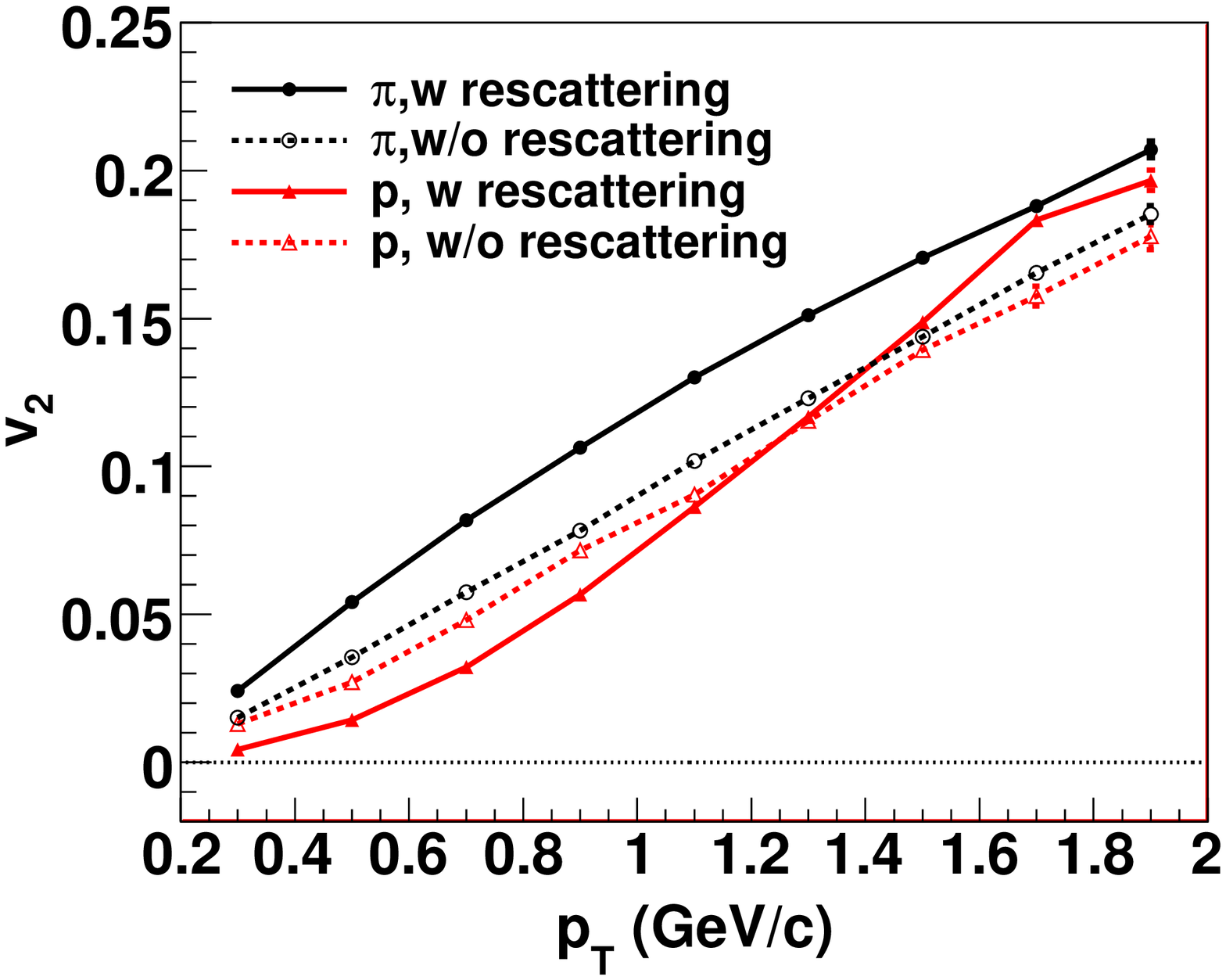}
\caption{(Left) $v_2(p_T)$ for identified hadrons. 
STAR data \cite{Adams:2004bi}
are compared to the hybrid model \cite{Hirano:2007ei}.
(Right) $v_2(p_T)$ with and without hadronic rescattering \cite{Hirano:2007ei}.
}
\label{fig:v2pt}
\end{figure}

The hybrid model also reproduces a mass ordering pattern of
$v_2$ for identified hadrons
as a function of $p_{T}$ near midrapidity \cite{Adams:2004bi}
(Fig.~\ref{fig:v2pt}, left) and in forward rapidity 
region \cite{Sanders:2007th}.
We consider semi-central collisions (20-30\% 
centrality), choosing impact parameter $b = 7.2$\,fm.
If one would look at the result just after the hadronisation,
the difference between pions and protons would be
quite small as shown in Fig.~\ref{fig:v2pt} (right).
As demonstrated in this figure, 
at low $p_T$ the pion curve moves up while the proton 
curve moves down through rescatterings.
So the hadronic rescatterings
are attributed to the observed
splitting patterns of $v_2(p_{T})$.
It should be noted that the same hadronic rescattering 
effect is also seen in the recent study based on
the different hadronic cascade model \cite{Werner}.
One can conclude that the large magnitude of the 
integrated $v_2$ and the strong mass ordering of the differential 
$v_{2}(p_{T})$ observed at RHIC result from a subtle interplay between 
perfect fluid dynamics of the early QGP stage and dissipative dynamics 
of the late hadronic stage.
The large magnitude of $v_2$ is due to 
the large overall momentum anisotropy, generated predominantly in the 
early QGP stage.
Whereas the strong mass splitting behavior
at low $p_T$ reflects the redistribution of this momentum 
anisotropy among the different hadron species during the 
hadronic rescattering phase.

\begin{figure}
\centering
\includegraphics[width=.46\linewidth]{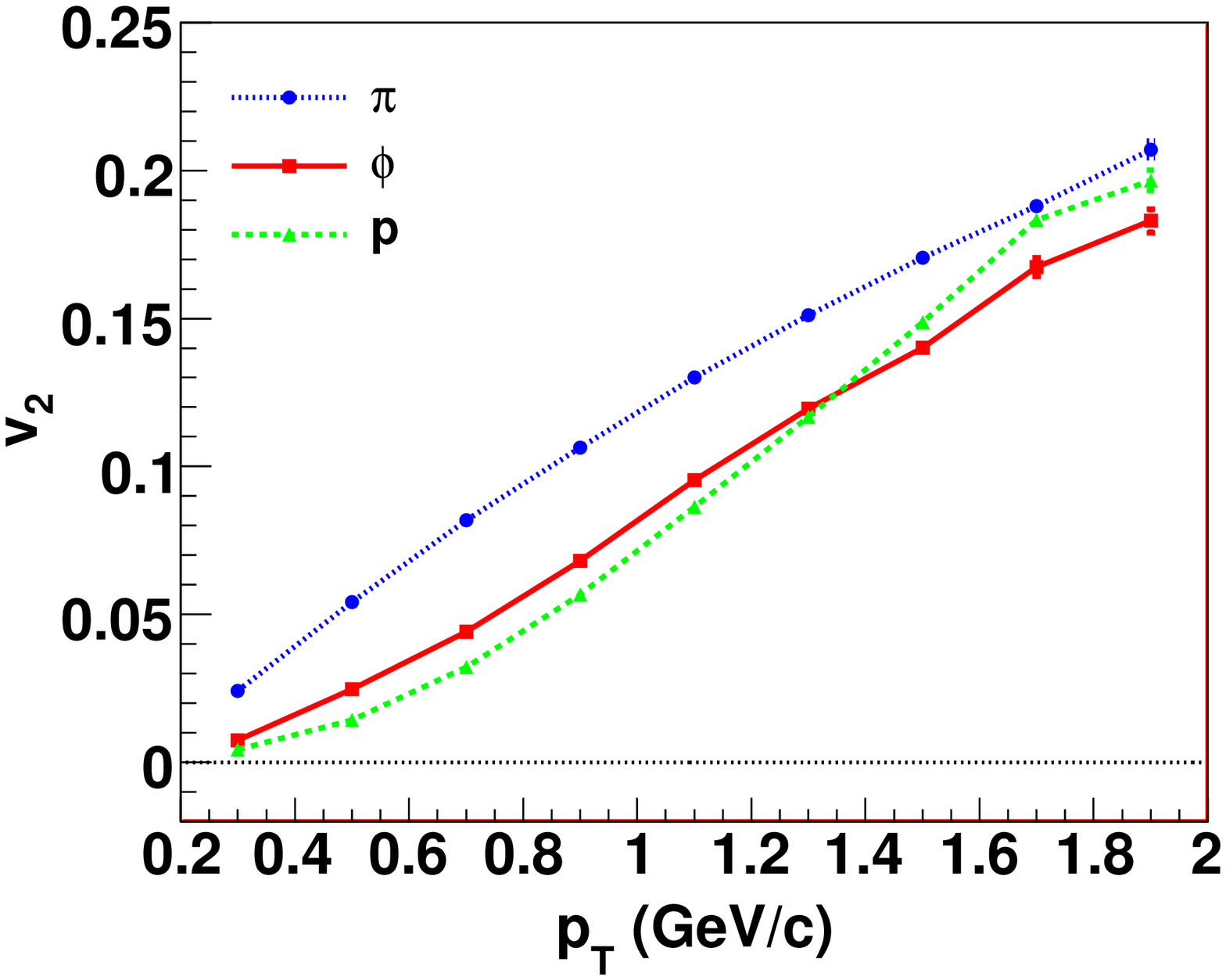}
\includegraphics[width=.46\linewidth]{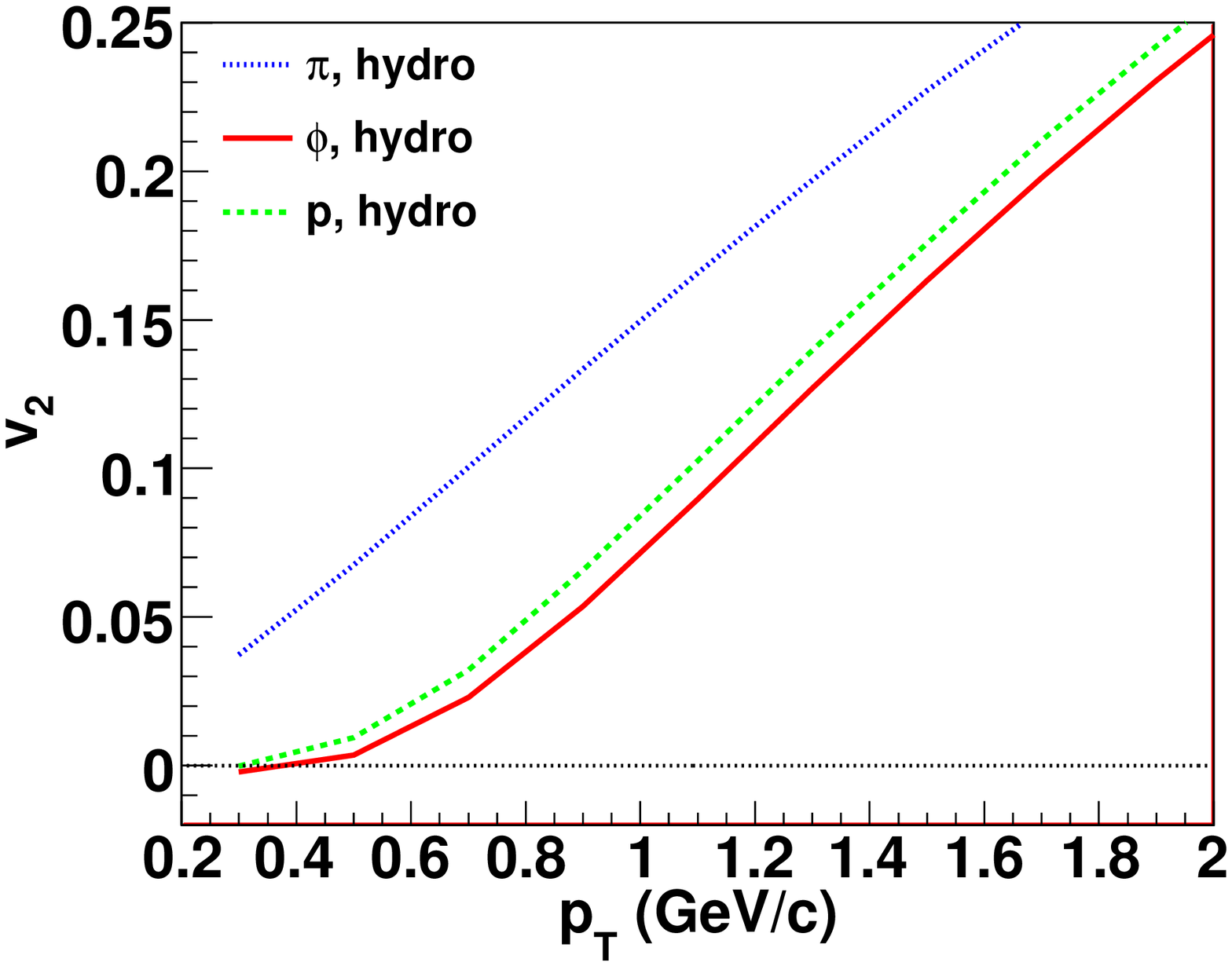}
\caption{$v_2(p_T)$ for pions (dotted), protons (dashed)
and $\phi$ mesons (solid) in Au+Au collisions at $b = 7.2$ fm
from the hybrid model (left) and from the ideal hydrodynamic model
with $T^{\mathrm{dec}}=100$ MeV (right).
}
\label{fig:v2pt_phi}
\end{figure}

Hadronic rescattering turns out to cause the mass ordering of
$v_2(p_T)$.
Then, what happens to the (hidden) strangeness sector
in which rescattering with 
other hadrons does not happen so frequently \cite{Shor:1984ui}? 
Figure \ref{fig:v2pt_phi} (left) shows $v_2(p_T)$ for 
$\pi$, $p$ and $\phi$.  
The $v_2(p_T)$ curves for pions and protons separate 
as discussed before,
whereas $v_2(p_T)$ for the $\phi$ meson remains almost unchanged
during the hadronic stage.
As a result of rescattering the proton elliptic 
flow ends up being smaller than that of the $\phi$ meson, 
$v_2^p(p_T){\,<\,}v_2^\phi(p_T)$ for $0{\,<\,}p_T{\,<\,}1.2$\,GeV/$c$, 
even though $m_\phi{\,>\,}m_p$. Hadronic dissipative effects are seen 
to be particle specific, depending on their scattering cross sections 
which couple them to the medium. The large difference 
of cross section 
between protons and $\phi$ mesons in the hadronic rescattering 
phase leads to a violation of the hydrodynamic mass ordering at low 
$p_T$ in the final state.
Note that clear mass ordering is seen
in Fig.~\ref{fig:v2pt_phi}
(right) if $\phi$ mesons were also assumed to fully participate
in hydrodynamic flow.

\section{Application of hydrodynamic results}

So far, we have seen that the hybrid model, in which 
the dynamical evolution of the locally thermalised QGP 
and the dissipative hadronic gas are incorporated,
works well in describing the elliptic flow data.
One can utilise the space-time evolution of the QGP obtained above
to study other observables.

The charmonium is one of the major diagnostic tools
of the QGP in relativistic heavy ion collisions \cite{MatsuiSatz}.
The melting temperature of $J/\psi$, $\psi'$ and $\chi_c$
is obtained by modeling the free streaming of
charmonia on the QGP background \cite{Gunji, Gunji2}. 
We first generate the charmonia
in the transverse plane, distribute them according to
the binary collision density,
simulate the transverse motion of charmonium and
then dissociate them when the temperature at the position
of a $J/\psi$ ($\psi'$ and $\chi_c$)
is larger than the melting temperature $T_{J/\psi}$ ($T_{\chi}$).
It is found \cite{Gunji, Gunji2}
that the suppression pattern as a function of centrality
reflects the region of the temperature above $T_{J/\psi}$
which is obtained in hydrodynamic calculations.
The best fit to
the suppression factor observed by PHENIX \cite{Adare:2006ns} is achieved  
by choosing $T_{J/\psi} = 2 T_c$ and $T_{\chi} = 1.34 T_c$.

Thermal photon is one of the direct and penetrating probes
to study the QGP \cite{Shuryak}. 
To calculate the transverse momentum spectra of thermal photons,
one convolutes the reaction rate, which is the yield per
unit space-time volume, over the volume of thermalised matter
in heavy ion collisions which is obtained from hydrodynamic
simulations.
By using the hydrodynamic results obtained in the previous section,
we calculate the thermal photon spectra \cite{FMLiu, FMLiu2}
as well as components of hard photons, 
jet-photon conversion and fragmentation.
We find that thermal photons originated from the QGP phase
are dominant around $p_{T} \sim 3$ GeV/$c$ \cite{FMLiu, FMLiu2}. 

From hydrodynamic simulations in the previous section,
the maximum temperature is found to be
at most $\sim$400 MeV in central collisions
which is much smaller than the mass of heavy quarks.
So it is safe to assume that these heavy quarks are
not thermally excited and are produced only in the
first hard collisions.
We treat these heavy particles as
``impurities" in the medium in which light degrees of freedom
such as $u$, $d$ and $s$ quarks and gluons are equilibrated.
We model the transverse motion of heavy quarks as the Brownian motion
and calculate the nuclear modification factor
for electrons and positrons from semileptonic decays of
heavy mesons
and elliptic flow parameter for heavy quarks.
We solve the following relativistic Langevin equations
\begin{eqnarray}
\Delta \vec{x}(t) & = & \frac{\vec{p}}{E(p)}\Delta t, \quad 
\Delta \vec{p}(t)  = -\Gamma \vec{p} \Delta t + \vec{\xi}(t)
\end{eqnarray}
on the hydrodynamic background \cite{Akamatsu}.
Here $M$, $\vec{x}$ and $\vec{p}$ are mass, position and momentum of
a sample heavy quark. $\vec{\xi}$ is the Gaussian white noise.
We parametrize drag force as $\Gamma = \gamma \frac{T^2}{M}$.
The above equations are solved in the local rest frame of a fluid element
which is moving with $u^{\mu}$ in the center of mass system.
\begin{figure}
\centering
\includegraphics[width=0.46\textwidth]{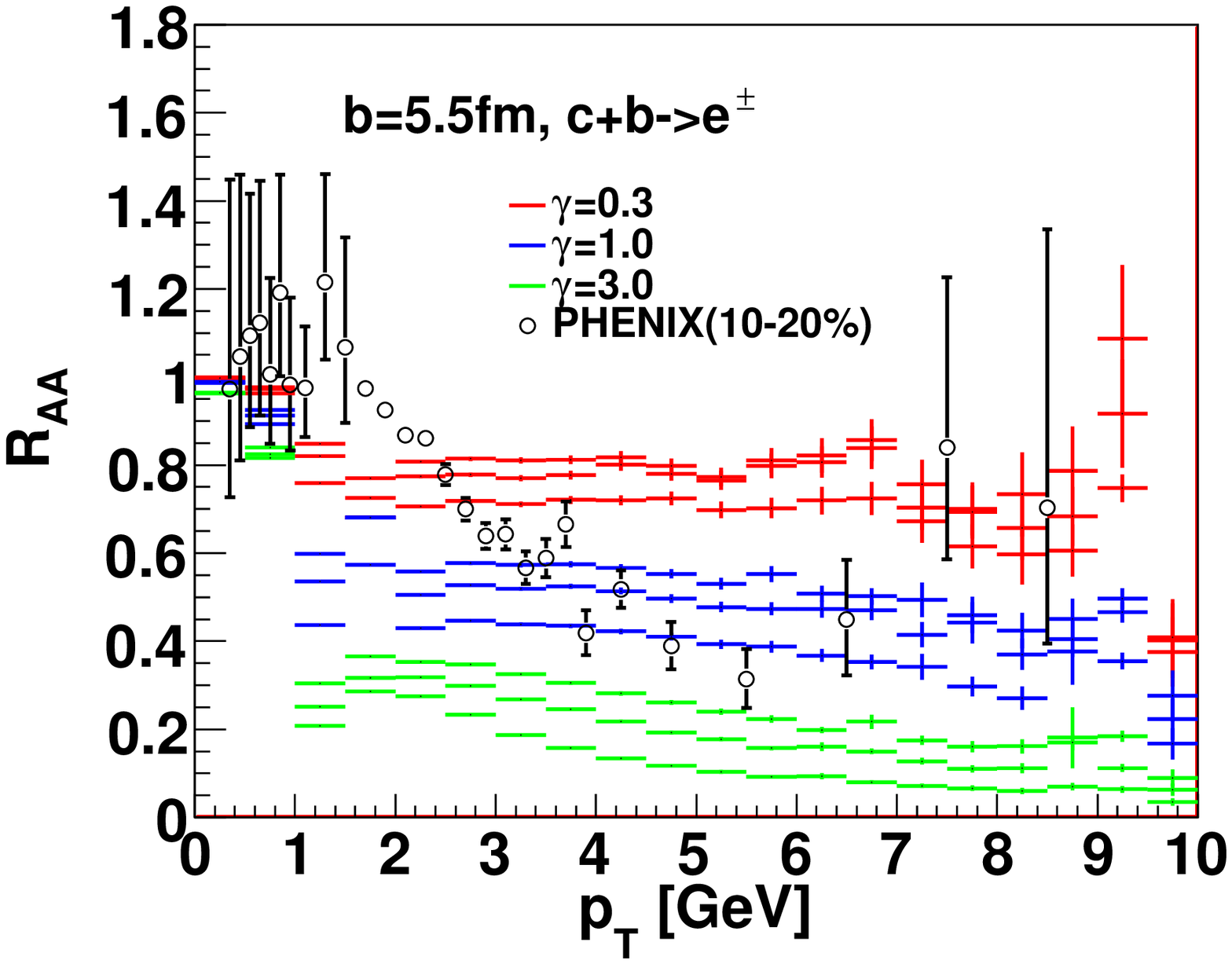}
\includegraphics[width=0.46\textwidth]{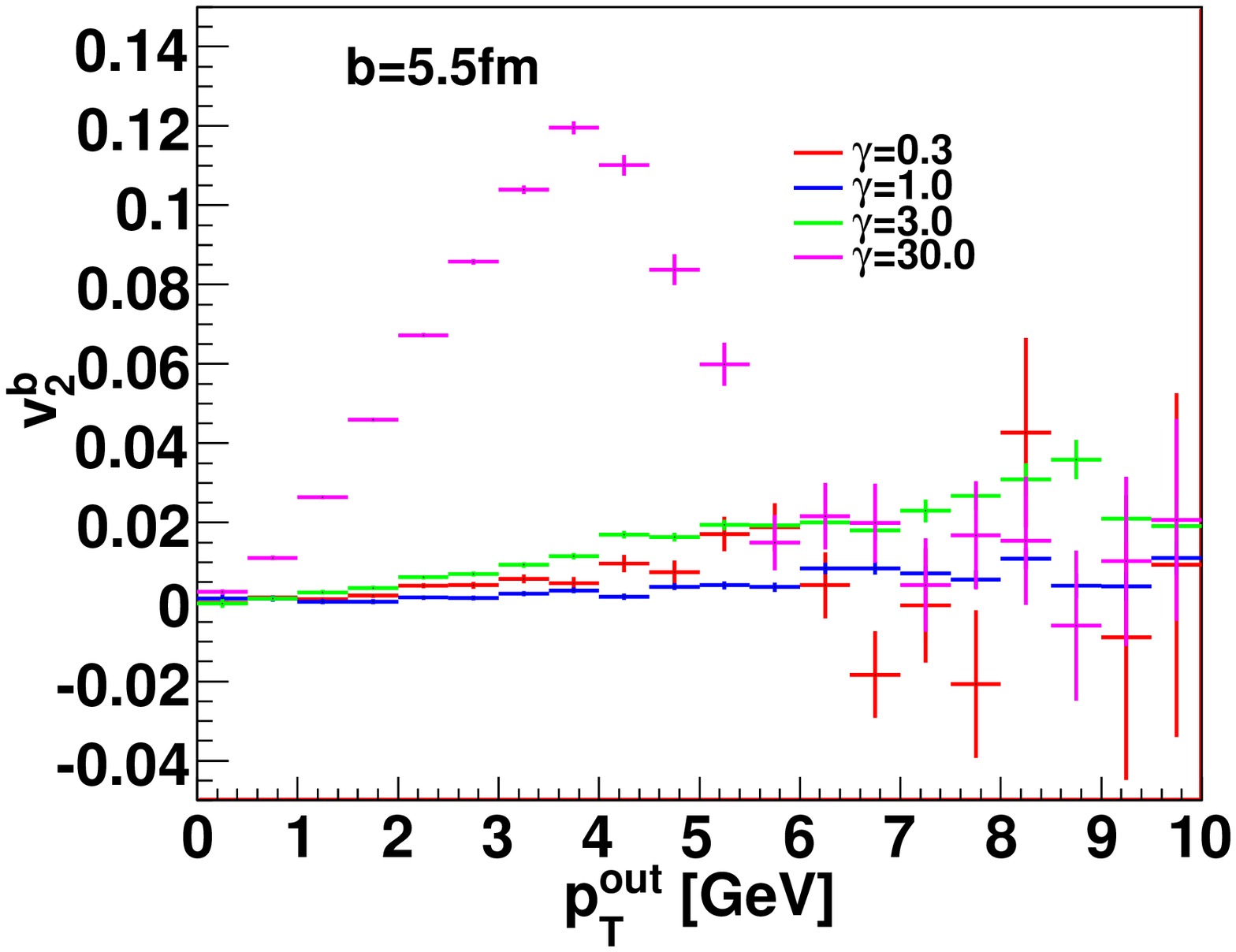}
\caption{(Left) Nuclear modification factor for electrons and positrons
from semileptonic decays in semi-central collisions
for $\gamma$ = 0.3, 1.0 and 3.0.
Experimental data are obtained by PHENIX \cite{Adare:2006nq}.
Upper and lower bands represent when diffusion of heavy quarks stops
in the mixed phase $f_{\mathrm{QGP}} = 1$ and 0, respectively,
where $f_{\mathrm{QGP}}$ is the fraction of the QGP phase in the
mixed phase. The middle points in the band
correspond to $f_{\mathrm{QGP}}$ = 0.5.
(Right) Sensitivity of $v_2(p_T)$ for bottom quarks to the parameter
in the drag force. Results are obtained with $f_{\mathrm{QGP}}$ = 0.5.}
\label{fig:v2pthq}
\end{figure}
$\gamma$ in the drag force is the only adjustable parameter
in this modeling of heavy quark diffusion.
From a comparison of nuclear modification factor 
with the data in Fig.~\ref{fig:v2pthq} (left),
we find $\gamma \sim$ 1-3 gives a reasonable description
of the data observed by PHENIX \cite{Adare:2006nq}.
If we look at $v_2(p_T)$
for heavy quarks just before hadronisation, the results with
$\gamma \sim$ 1-3 are considerably smaller than the result 
with $\gamma=30.0$ whose relaxation time $\tau = 1/\Gamma$ is
much smaller than the life time of the QGP.
This indicates that the heavy quarks
do not reach the complete thermalisation limit.

\section{Summary}

We discussed the current status of hydrodynamic description
of relativistic heavy ion collisions.
Integrated and differential elliptic flow parameters
in low $p_T$ regions
are well reproduced by the hybrid approach
in which ideal hydrodynamic description
of the QGP is followed by microscopic description
of hadron gas. 
We found the observed mass ordering pattern
results from the hadronic rescatterings.
As a consequence, violation of mass ordering is predicted
for $\phi$ mesons that do not couple to the hadronic
medium.
We also discussed the recent application of hydrodynamic results
to $J/\psi$ suppression, thermal photon radiation and
heavy quark diffusion.


\ack
The author would like to thank
Y Akamatsu, T Gunji, H Hamagaki, 
T Hatsuda, U Heinz, D Kharzeev, R Lacey,
F M Liu, Y Nara, K Werner and Y Zhu
for fruitful discussion.
The work was partly supported by Grant-in-Aid for Scientific
Research No.~19740130 
and by Sumitomo Foundation No.~080734.

\end{document}